\documentclass[runningheads]{llncs}
\usepackage{amsmath}
\usepackage{booktabs}

\usepackage{tikz}
\usepackage[T1]{fontenc}

\usepackage{subcaption} 
\usepackage{multirow}
\usepackage{graphicx}

\begin{document}

\title{Evaluating Explanation Quality in X-IDS Using Feature Alignment Metrics}

\author{ Mohammed Alquliti\inst{1} \and
Erisa Karafili\inst{1} \and
BooJoong Kang\inst{1}
}
\authorrunning{M. Alquliti et al.}

\titlerunning{Evaluating Explanation Quality in X-IDS}
\institute{
Electronics and Computer Science Department, University of Southampton, Southampton, United Kingdom \\
\email{\{M.H.Alquliti, E.Karafili, B.Kang\}@soton.ac.uk}
}

\maketitle             

\begin{abstract}
Explainable artificial intelligence (XAI) methods have become increasingly important in the context of explainable intrusion detection systems (X-IDSs) for improving the interpretability and trustworthiness of X-IDSs. However, existing evaluation approaches for XAI focus on model-specific properties such as fidelity and simplicity, and neglect whether the explanation content is meaningful or useful within the application domain. In this paper, we introduce new evaluation metrics measuring the quality of explanations from X-IDSs. The metrics aim at quantifying how well explanations are aligned with predefined feature sets that can be identified from domain-specific knowledge bases. Such alignment with these knowledge bases enables explanations to reflect domain knowledge and enables meaningful and actionable insights for security analysts. In our evaluation, we demonstrate the use of the proposed metrics to evaluate the quality of explanations from X-IDSs. The experimental results show that the proposed metrics can offer meaningful differences in explanation quality across X-IDSs and attack types, and assess how well X-IDS explanations reflect known domain knowledge. The findings of the proposed metrics provide actionable insights for security analysts to improve the interpretability of X-IDS in practical settings.

\end{abstract}

\section{Introduction}

Nowadays, cyber threats continuously increase in complexity to evade intrusion detection systems (IDSs). This has led to the use of machine learning-based intrusion detection systems (ML-IDSs). ML-IDSs have shown potential in improving detection capabilities by analysing and adapting to complex patterns and anomalies in network traffic \cite{capuano2022explainable}. Despite their advantages, ML-IDSs often fail to provide the reasoning behind decisions (e.g., why detected activities are suspicious). This leaves security analysts, consequently, also subsequent response mechanisms, with insufficient information about detected suspicious activities. This problem requires explainable IDSs (X-IDSs) utilising explainable AI (XAI) method(s) \cite{moustafa2023explainable}, such as local interpretable model-agnostic explanations (LIME) \cite{dieber2020} and shapley additive explanations (SHAP) \cite{lundberg2017unified}. In practice, such post hoc XAI methods can highlight which network flow features have most influenced the prediction of ML-IDSs. 

As XAI methods are increasingly applied to X-IDSs, evaluating the quality of explanations becomes critical, as analysts rely on clear, understandable explanations to make decisions effectively \cite{neupane2022}. There have been XAI evaluation methods focusing on generic properties, such as how well it mimics the model's behaviour (fidelity) \cite{arya2019}, or how concise it is (simplicity) \cite{lopes2022}. However, there has been a lack of research on developing evaluation methods focusing on the quality of explanations in the domain-specific context. As a result, current evaluation methods do not account for whether an explanation aligns with domain expectations or highlights indicators of known attack patterns \cite{pawlicki2024,neupane2022}.

To address this gap, we propose new evaluation metrics that assess the quality of explanations in the cyber security context, more specifically, X-IDSs. The aim of the proposed metrics is to quantitatively evaluate how well explanations capture domain-informed features that support the analyst's understanding of the detected attack. They are based on comparing the top-\(k\) features identified in an explanation with a predefined set of domain-informed features that are considered important to detect a specific type of attack.  These sets are used by our metrics to assess whether the X-IDS is highlighting the features that matter most from a cyber security perspective. The proposed metrics are: Feature Alignment Precision (FAP), Feature Alignment Recall (FAR), and Feature Alignment F1 (FAF1). In essence, these metrics measure the alignment between the explanation of an X-IDS and the domain-informed features of an attack derived from cybersecurity knowledge domain resources. The FAP measures the proportion of an explanation top features that are relevant according to domain knowledge. The FAR measures the proportion of the domain-informed features that the explanation of an X-IDS managed to capture. The FAF1 provides a balanced single measure of explanation quality in terms of domain alignment. 

To evaluate these metrics, we applied them to explanations generated by three X-IDSs: Random Forest (RF)~\cite{arreche2024xai}, Deep Neural Network (DNN)~\cite{arreche2024xai}, and CNN-BiLSTM~\cite{sinha2020efficient}. All models were trained on the CICIDS2017 benchmark intrusion detection dataset~\cite{sharafaldin2018}. The goal of the evaluation was to determine how well each explanation aligned with domain-informed feature sets across different types of attacks. We assessed alignment at three levels of evaluation: instance level (individual explanations), attack class level (aggregated by attack type), and dataset level (overall performance across all examples). Our findings show that explanation quality varies across models and attack types. For example, the DNN and CNN-BiLSTM achieved higher FAP and FAR at lower values of \(k\). This suggests that these X-IDSs are more effective at identifying most of the domain-informed features earlier in the explanation process. These results show that the proposed metrics can effectively distinguish explanation quality across X-IDSs and assess whether an X-IDS highlights the features that matter most for the detection.

In the remainder of this paper, we start with an overview of the related work on explainable IDS in Section 2. In Section 3, we present our methodology, where each metric is introduced, together with how it is calculated and evaluated. In Section 4, we present our experimental setup and results to illustrate how our metrics reveal differences in explanation quality across various X-IDSs. We conclude the paper and discuss future potential research directions in Section 5.

\section{Related Work}
In this section, existing research is reviewed in three areas that are foundational to our work. First, X-IDS is examined while focusing on the nature of the explanations they provide and the importance of interpretability for end-users such as security analysts. Next, the use of domain knowledge frameworks like MITRE ATT\&CK and D3FEND to contextualise security tasks. Finally, we summarise recent approaches for evaluating explanation quality in XAI, and highlight the lack of metrics that explicitly consider alignment with domain-informed knowledge.

\subsection{Explainable Intrusion Detection Systems}

Recently, a growing emphasis has been placed on the need for XAI methods to provide explanations tailored to end-users, rather than solely interpretable by developers and researchers. In \cite{moustafa2023explainable}, authors argue that many XAI methods produce low-level explanations, typically in the form of numerical feature importance vectors. These low-level explanations are useful for developers and researchers to understand the internal behaviour of models. However, they lack the contextual interpretation needed for end users (e.g., security analysts).  In contrast, high-level explanations aim to relate model outputs to broader security concepts, such as known attack tactics or behaviours, making them more accessible to security analysts. Such explanations provide the contextual clarity and actionable insights that security analysts need. For example, 
\cite{lanvin2023towards} presents Auto-Encoder (AE)-pvalues, an explanation method for unsupervised network intrusion detection systems that identifies abnormal network traffic using autoencoder-based anomaly scores. However, the explanations remain low-level as they highlight numerical deviations in features without indicating their operational relevance. The authors of \cite{lanvin2023towards} indicate that this limitation is derived from factors such as high feature correlations, dataset biases, and the model's focus on individual network connections without contextual information about expected values. Similarly, \cite{hasan2023explainable} presents an explainable Deep Learning (DL)-based IDS aimed at enhancing the transparency and robustness of DL-based IDSs. This solution applies SHAP and LIME techniques to generate low-level explanations in the form of numerical feature importance vectors, supporting analysts in the following steps or processes. As these studies~\cite{moustafa2023explainable,lanvin2023towards,hasan2023explainable} highlight that low-level explanations may not be meaningful to security analysts, this limitation highlights the lack of alignment between explanations and the domain-specific knowledge that analysts rely on during investigations. To address this, \cite{hasan2023explainable} emphasised the necessity of collaborative efforts in advancing XAI to communicate results effectively to non-AI experts. Furthermore, incorporating domain-specific knowledge into XAI methods is crucial for improving model interpretability and elevating explanations to high-level and actionable insights \cite{wu2023black}. Additionally, based on \cite{wu2023black}, it is essential to go beyond mere identification of feature importance and pursue conceptual-level explanations by incorporating domain-specific knowledge. Existing work demonstrates the need for explanations to reflect such domain knowledge. Our work builds on this foundation by introducing metrics to evaluate how well explanations, produced by SHAP, align with domain-specific knowledge.

\subsection{Domain Knowledge Frameworks} 

As high-level explanations are needed, standardised domain-specific knowledge bases to encode domain expertise are increasingly leveraged. For example, MITRE ATT\&CK can help to contextualise the IDS outputs. The MITRE ATT\&CK matrix serves as a repository of tactics and techniques \cite{mitre_attack,cisa2023}. Mapping the IDS detections to these known tactics and techniques can make explanations more actionable for security analysts. Arreche et al. \cite{arreche2024xai} recently demonstrated this approach by referencing network attack classes with relevant ATT\&CK tactic and technique for IDs. For instance, Denial of Service (DoS) attack in CICIDS2017 dataset can be labelled as Network Denial of Service [MITRE ATT\&CK ID: T1498]. This alignment allows an IDS to explain alerts in terms of the existing offensive techniques in MITRE. This approach can bridge the gap between low-level features and high-level detection explanations. Similarly, the idea of linking IDS detections to MITRE ATT\&CK is becoming popular, e.g., Daniel et al. \cite{nir2025labeling} uses automation to label network IDS signatures with the appropriate MITRE ATT\&CK tactics and techniques. This approach ensures that any alerts generated by these rules include an explanation of the adversarial technique. This solution helps incident responders manage the triaging phase. Additionally, MITRE D3FEND offers a complementary knowledge base of defensive techniques and countermeasures \cite{kaloroumakis2021}. D3FEND provides defensive actions and links them to the ATT\&CK techniques they mitigate. In practice, they enable defining sets of domain-informed features or indicators based on known domain-specific knowledge. Our proposed explanation evaluation metrics assess how well the most influential features identified by an X-IDS correspond to these domain-informed features.

\subsection{Evaluation of Explanations}

Evaluating the quality of explanations from XAI methods has been addressed through qualitative approaches, typically through human-centred studies \cite{schwalbe2024,zhou2021}. However, many surveys note that formal evaluation for explanation quality is frequently assumed or judged rather than measured \cite{schwalbe2024,zhou2021,nauta2023}. Despite the lack of generalised accepted metrics to evaluate XAI approaches, qualitative metrics have the potential to ultimately establish a standardised and quantified means of evaluation \cite{rosenfeld2021,hedstrom202X,schwalbe2024}. In \cite{hedstrom2023}, the authors provided a set of five distinct properties for evaluation: \emph{faithfulness, robustness, localisation, randomisation}, and \emph{complexity}. Properties like faithfulness (e.g., how the model's influencing features truly affect its explanation changes) and robustness (e.g., how stable a model's explanation remains with minimal changes to input) focus merely on the XAI method behaviour. In IDS context, \cite{arreche2024xai} presented an end-to-end framework for evaluating both global and local explanations using SHAP and LIME, and defines six metrics for explanation quality, i.e., descriptive accuracy/fidelity, efficiency, stability, sparsity, robustness, and completeness. The metrics used in \cite{lin2021} are closer to our work as they compare the identified features of the XAI method to a ground truth of important features. \cite{lin2021} provides a score similar to precision-recall, but lacks the division between correctness and completeness. 

The above metrics do not explicitly verify whether explanations align with domain-specific knowledge, particularly in the IDS context. To address this gap, we introduce formally defined metrics, namely \emph{FAP}, \emph{FAR}, and \emph{FAF1}. In our related experiments, we compare outputs of the explanation method using these metrics and analyse the alignment at various levels (per instance, per attack class, and across the entire dataset).

\section{Methodology}

\begin{figure}
    \centering
    \includegraphics[width=.90\linewidth]{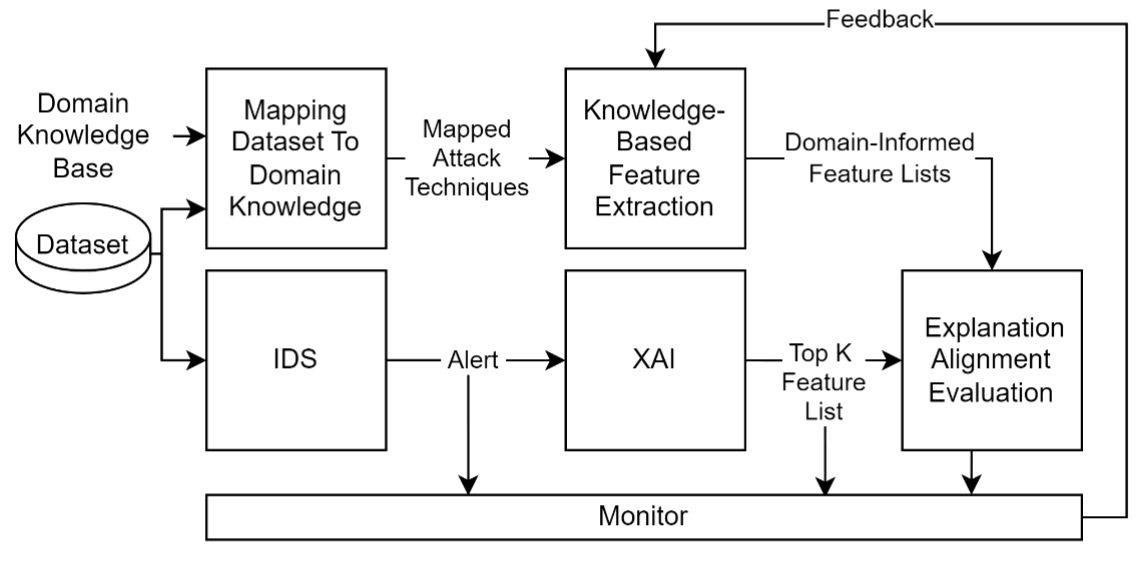}
    \caption{High-level overview of the explanation evaluation process. The top‑\(k\) features from the XAI model are evaluated against domain-informed feature sets. } 
    \label{fig:High-level methodology}
\end{figure}

This section introduces the proposed evaluation metrics that assess the quality of explanations produced by X-IDSs. These X-IDSs utilise ML/DL-based IDS models with post hoc explanation techniques, such as LIME and SHAP, which identify the reasoning behind a model's predictions by highlighting the most influential features, as illustrated in Figure \ref{fig:High-level methodology}. We check how well the most influential features align with domain-specific knowledge. Our proposed metrics evaluate this alignment by comparing the explanations of the model against predefined sets of domain-informed features for specific attack classes.

To use our evaluation metrics, we require predefined sets of domain-informed features for each attack class. These feature sets capture domain knowledge and can be derived from domain-specific knowledge bases, expert knowledge, or security frameworks. In this work, we use feature sets derived from the MITRE ATT\&CK and D3FEND frameworks as a representative example. The procedure used to construct these sets is described in the experimental setup (Section~\ref{sec:experimental-setup}).

\subsection{System Model}

This section formalises the structure of an X-IDS used in our work by defining the components involved in generating and evaluating explanations, including the dataset, prediction model, explanation method, and the reference feature sets used to assess explanation relevance. 

Let \( \mathcal{D} \) denote the dataset used for training an ML-IDS. Each instance \(d \in \mathcal{D}\) is defined as \( d = \{f_1, f_2, \ldots, f_n, c\} \) where \(f_1, f_2, \ldots, f_n\) are the features of the instance (e.g., the characteristics of a network traffic record), and \( c \) is the class assigned to the instance (e.g., a specific attack type or benign).  Here, we denote the feature vector of an instance \(d\) as \(x = \{f_1, f_2, \ldots, f_n\}\), and let \( \mathcal{X}\) be the set of all feature vectors in the dataset and let \( \mathcal{C} \) be the set of all corresponding classes. For training and evaluation, let \( \mathcal{X}^{\text{train}} \subseteq \mathcal{X} \) and \( \mathcal{X}^{\text{test}} \subseteq \mathcal{X} \) denote the disjoint training and testing sets, respectively. The corresponding classes denoted as \( \mathcal{C}^{\text{train}} \subseteq \mathcal{C} \) and \( \mathcal{C}^{\text{test}} \subseteq \mathcal{C} \).

A classifier \(f\) learns a mapping from feature vectors to corresponding classes, denoted as \(f: \mathcal{X} \rightarrow \mathcal{C}\). A classifier \(f\) is trained on the feature vectors in \( \mathcal{X}^{\text{train}}\) and corresponding classes in \( \mathcal{C}^{\text{train}}\). Once trained, the classifier produces predictions \( f(x) = \hat{c} \)  for the feature vector of an instance \( x \in \mathcal{X}^{\text{test}} \), where \(\hat{c}\) is the predicted class. Also, let \( \mathcal{C}^\text{attack} \subseteq \mathcal{C} \) denote the set of all attack classes considered for evaluation, excluding the benign class.

To make the prediction \(\hat{c}\) interpretable, an XAI method is applied. An XAI method \( g(f, x) = E_x\) takes a trained classifier \( f \) and a feature vector of an instance \( x \in \mathcal{X} \) and returns an ordered set of features \(E_x\) that most influenced the classifier’s prediction for instance \(x\). The features in \(E_x\) are ordered by their importance scores, which reflect how much each feature contributed to the model’s decision. An X-IDS often select the top-\(k\) features from \(E_x\) denoted as \(E_x(k)\) where \( E_x(k) \subseteq E_x \).

\subsection{Explanation Evaluation Metrics}
\label{sec:metrics}

This section introduces our evaluation metrics used to evaluate the quality of the explanations produced by X-IDSs.
The evaluation metrics are: Feature Alignment Precision (FAP), Feature Alignment Recall (FAR), and Feature Alignment F1 (FAF1). They quantify how well explanations of an X-IDS correspond to the domain-informed feature sets defined by security domain knowledge. The FAP metric measures the fraction of the top‑\(k\) most influential features that are relevant to the sets of domain-informed features defined by security domain knowledge. In other words, it counts how many of the top-\(k\) features from the explanation are present in predefined sets of domain-informed features. The FAR metric measures the fraction of the domain-informed features that are present in the top‑\(k\) features produced by the X-IDS. It measures how well the explanation covers the critical indicators identified by security domain knowledge. The FAF1 metric offers a balanced overall score that captures both the correctness and the completeness of the explanations. 

In evaluating the quality of the explanations, the set of top-\(k\) features \( E_x(k) \) is compared against a set of domain-informed features associated with the predicted class. Let \( F_c \) denote the set of domain-informed features that are associated with a class \( c \in \mathcal{C}^\text{attack} \). These feature sets need to be predefined and can be derived from structured cyber security knowledge bases, expert input, or other relevant examples. These predefined sets are used as a reference to assess how well the explanation \( E_x(k) \) aligns with domain knowledge. The comparison between \( E_x(k) \) and \( F_c \) enables the computation of our proposed evaluation metrics.

Each metric can be evaluated at three different levels: dataset, class, and instance levels.  First is the dataset level, which uses all instances in the test set, denoted by \(\mathcal{X}^\mathrm{test}\). Second is the class level, which considers subsets of test instances that share the same label \(c\), denoted as \(\mathcal{X}_\mathrm{c}^\mathrm{test} \subseteq \mathcal{X}^\mathrm{test}\). Lastly, the instance level evaluation focuses on a single instance \(x\).

Our methodology evaluates the quality of \(E_x(k)\), the explanation for instance \(x\), by computing against \(F_c\), the set of domain-informed features for the true class \(c\). This comparison enables us to assess how well the model’s explanation aligns with domain-informed indicators. 
Below, we formally define the key concepts and present how each of our evaluation metrics is calculated. 

\subsection*{Feature Alignment Precision}
Feature Alignment Precision (FAP) measures the correctness of the explanation of an X-IDS according to a set of domain-informed features. It quantifies how many of the top-\(k\) features selected by the explanation method are also present in the reference set of domain-informed features for a given attack class. We define this metric at the instance, class, and dataset levels.

The \emph{instance-level FAP} captures how well the explanation for an individual feature vector of an instance \( x \) aligns with the domain-informed features expected for its true class \( c \). It is defined as:

\begin{equation}
\label{eq:instance_precision}
\text{FAP}_I (x,k)\;=\; \frac{\lvert\,E_x(k) \,\cap\, F_c\rvert}{\lvert E_x(k)\rvert}
\end{equation}

where \( E_x(k) \) is the set of top-\(k\) features produced by the explanation method for the feature vector of an instance \( x \), and \( F_c \) is the set of domain-informed features corresponding to class \( c \). A higher \( \text{FAP}_I(x,k) \) value indicates that a greater proportion of the selected features are relevant, which suggest better explanation quality for that instance.

The \emph{class-level FAP} evaluates explanation correctness within a specific class label for more detailed scores. It is computed as the average instance-level FAP across all test samples belonging to a class \( c \):

\begin{equation}
\label{eq:class_metrics}
\text{FAP}_{\mathrm{C}}(c, k) \;=\;
\frac{1}{|\mathcal{X}_c^{\mathrm{test}}|} \sum_{x \in \mathcal{X}_c^{\mathrm{test}}} \text{FAP}_I(x,k)
\end{equation}
where \(k\) is the specified number of top features, \( \mathcal{X}_c^{\mathrm{test}} \) denotes the set of test instances that are correctly predicted as the class \( c \), and \( \text{FAP}_I(x,k) \) is the instance-level FAP. 

The \emph{dataset-level FAP} provides a high-level overview of the explanation quality across all attack classes for the evaluated X-IDS. This FAP value provides a single overall number across the entire dataset’s performance, excluding benign traffic. It is computed as the average of class-level FAP scores over all classes used in the evaluation:

\begin{equation}
\label{eq:dataset_metrics_P}
\text{FAP}_{\mathrm{D}}(k) \;=\;
\frac{1}{|\mathcal{C}^\text{attack}|} \sum_{c \in \mathcal{C}^\text{attack}} \text{FAP}_{\mathrm{C}}(c, k)
\end{equation}
where \(k\) is the specified number of top features, \( \mathcal{C}^\text{attack} \) denotes the set of classes considered for evaluation. Each \( \text{FAP}_{\mathrm{C}}(c, k) \) is the average precision for class \( c \).

\subsection*{Feature Alignment Recall}

Feature Alignment Recall (FAR) measures how many of a set of domain-informed features are captured in the produced explanations. It reflects the completeness of the produced top-\(k\) in covering the set of domain-informed features.  Similarly to FAP, FAR is computed at the instance, class, and dataset levels. 

The \emph{instance-level FAR} for the feature vector of an instance \( x \) with the correctly predicted as the class \( c \) is computed as:

\begin{equation}
\label{eq:instance_recall}
\text{FAR}_I(x,k) = \frac{|E_x(k) \cap F_c|}{|F_c|}
\end{equation}
where \( E_x(k) \) is the set of top‑\(k\) features produced by the XAI method for the feature vector of the instance \( x \), and \( F_c \) is the set of domain-informed features for true class \( c \). A higher \( \text{FAR}_I(x,k) \) value indicates better completeness of the explanation for that instance.

The \emph{class-level FAR} is calculated by averaging the instance-level \( \text{FAR}_I(x,k) \) over all test instances that belong to a specific class \( c \):

\begin{equation}
\label{eq:class_metrics_R}
\text{FAR}_{\mathrm{C}}(c, k) \;=\;
\frac{1}{|\mathcal{X}_c^{\mathrm{test}}|} \sum_{x \in \mathcal{X}_c^{\mathrm{test}}} \text{FAR}_I(x,k)
\end{equation}
where \( \mathcal{X}_c^{\mathrm{test}} \) is the subset of test instances that are correctly predicted as the class \( c \). 

The \emph{dataset-level FAR} provides a high-level overview of how well explanations cover the domain-informed feature sets across all attack classes. It is computed as the average of class-level FAR scores as follows:

\begin{equation}
\label{eq:dataset_metrics_R}
\text{FAR}_{\mathrm{D}}(k) \;=\;
\frac{1}{|\mathcal{C}^\text{attack}|} \sum_{c \in \mathcal{C}^\text{attack}} \text{FAR}_{\mathrm{C}}(c, k)
\end{equation}
where \(k\) is the specified number of top features, and \( \mathcal{C}^\text{attack} \) is the set of classes included in the evaluation.

\subsection*{Feature Alignment F1}

Similar to the traditional F1-score in classification, the Feature Alignment F1 (FAF1) reflects a harmonic mean of \(FAP\) and \(FAR\) into a single score. FAF1 provides a balanced measure that captures both the correctness and the completeness of the top‑\(k\) explanations. We define this metric at the instance, class, and dataset levels.

The \emph{instance-level FAF1} is computed as the harmonic mean of instance-level FAP and FAR for each feature vector of instance \( x \):

\begin{equation}
\label{eq:balanced_score_x}
\text{FAF1}_I(x,k) = \frac{2 \cdot |E_x(k) \cap F_c|}{|E_x(k)| + |F_c|}
\end{equation}
where \( E_x(k) \) is the set of top‑\(k\) features produced by the XAI method for the feature vector of instance \( x \), and \( F_c \) is the domain-informed feature set for its class \( c \). A high \( \text{FAF1}_I(x,k) \) score indicates that the explanation captures a greater proportion of the domain-informed feature set while minimising the inclusion of irrelevant ones.

The \emph{class-level FAF1} is defined as the average instance-level FAF1 across all test instances with class \( c \) defined below.

\begin{equation}
\label{eq:balanced_score_c}
\text{FAF1}_{\mathrm{C}}(c, k) = \frac{1}{|\mathcal{X}_c^{\mathrm{test}}|} \sum_{x \in \mathcal{X}_c^{\mathrm{test}}} \text{FAF1}_I(x,k)
\end{equation}

The \emph{dataset-level FAF1} aggregates the class-level FAF1 scores over all attack classes as follows:

\begin{equation}
\label{eq:balanced_score_d}
\text{FAF1}_{\mathrm{D}}(k) \;=\;
\frac{1}{|\mathcal{C}^\text{attack}|} \sum_{c \in \mathcal{C}^\text{attack}} \text{FAF1}_{\mathrm{C}}(c, k)
\end{equation}
where \(k\) is the specified number of top features, \( \mathcal{C}^\text{attack} \) denotes the set of evaluated classes, and \( \text{FAF1}_{\mathrm{C}}(y, k) \) is the class-level FAF1 for class \( c \).

\subsection{Perspectives of our Explanation Evaluation Metrics}
Although we use the terms ``precision'', ``recall'' and ``F1'' in our metrics  (e.g., Equations \ref{eq:instance_precision}-\ref{eq:balanced_score_d}), their interpretation differs from the conventional precision, recall, and F1 that are used in standard classification tasks. In standard classification, precision measures the proportion of correctly classified instances among all predicted positives, while recall measures the proportion of correctly classified instances among all true positives.
In contrast, our \emph{explanation evaluation} metrics assess the quality of explanations by comparing the top-\(k\) most influential features \(E_x\) produced by an XAI method to a set of domain-informed features \(F_c\). Rather than evaluating correctness at the instance level, our metrics evaluate correctness and completeness at the feature level. In other words, our metrics focus on whether the explanation highlights features that align with the domain-informed feature set. A high FAP means that most of the features selected by the model belong to the set of domain-informed features \(E_x\), while a high FAR means we are capturing most of the domain-informed features that matter. 

\section{Evaluation}

The evaluation aims to measure how well explanations generated by X-IDSs align with domain-specific knowledge bases in the intrusion detection context. In this section, we present the details of our evaluation process and results. In the experimental setup, we introduce the derivation of the domain-informed feature sets using the MITRE ATT\&CK and D3FEND frameworks \cite{cisa2023,kaloroumakis2021}. We continue by explaining the dataset, ML-IDS, and XAI method used in the evaluation. Finally, we analyse the experimental results from comparing the most influential features of each model against the domain-informed features sets using our evaluation metrics at different levels.

\subsection{Experimental Setup}\label{sec:experimental-setup}

Our evaluation includes two key components: (1) generating domain-informed feature sets that reflect what should be relevant for each attack type, and (2) extracting the top-\(k\) features from the explanations that was generated by X-IDS to compare and assess them against the domain-informed feature sets.

To construct the domain-informed feature sets, we mapped each attack class in the dataset to its corresponding offensive techniques in the MITRE ATT\&CK framework. The identified techniques were linked to defensive tactics in MITRE D3FEND to derive domain-informed feature sets. In particular, we focused on the detect tactics, which outline specific indicators that security analysts can use to identify malicious behaviour. Also, we further enriched these sets by consulting contextual ATT\&CK resources, such as detection recommendations, mitigation strategies, and real-world examples, to enhance the feature extraction of each attack. The derived feature sets were used as the domain-specific knowledge to evaluate the quality of the explanations generated by the X-IDS. It is important to note that while we rely on this process to produce domain-informed feature sets, the primary aim of this study is not to propose a new derivation method, but rather to evaluate the effectiveness of our explanation metrics using these sets as a baseline.

To evaluate the metrics, our experiments are based on the CICIDS2017 benchmark intrusion detection dataset \cite{sharafaldin2018}. A balanced subset was extracted using undersampling \cite{kostas2018} to mitigate the impact of class imbalance. This preprocessed dataset was used to train three ML/DL-IDSs: a Random Forest (RF) and a Deep Neural Network (DNN), both adapted from \cite{arreche2024xai}, and a hybrid CNN-BiLSTM architecture \cite{sinha2020efficient}. RF was selected as a traditional, interpretable model that performs robustly and can be easily explained using feature-based methods, while DNN was chosen as a strong deep learning baseline that offers higher detection accuracy but requires post hoc explanation due to its complexity. The CNN-BiLSTM model was included as it combines convolutional and bidirectional LSTM layers to capture both spatial and temporal patterns in network traffic. These models demonstrated robust and consistent detection capabilities across attack classes in their original studies \cite{arreche2024xai,sinha2020efficient}.

All three used ML/DL-IDS are followed by an XAI method, SHAP, for the explainability part \cite{arreche2024xai}. SHAP is a post hoc model-agnostic ML explainability approach that can be used after the AI models and assigns to each feature a score that represents the feature contribution to the reached prediction \cite{lundberg2017unified}. We used the generated explanations against the domain-informed feature sets using the proposed explanation evaluation metrics presented in Section~\ref{sec:metrics}. 

\subsection{Experimental Results}
The performance of the explanation evaluation metrics is evaluated on three X-IDS (RF and DNN \cite{arreche2024xai}, and CNN-BiLSTM \cite{sinha2020efficient}). We assess how well their explanations align with domain-informed features derived from MITRE frameworks. In particular, we generated SHAP explanations for each model's prediction on the test set and compared the top-\(k\) features of the explanations to the MITRE-based feature sets for the corresponding attack class. The evaluation is performed at multiple levels (instance level, class level, and dataset level) to provide a comprehensive view of the explanations' quality. We introduce in Table~\ref{tab:dos_cleaned} and Figure~\ref{fig:Dataset_level_plot} each model’s overall alignment performance on the full dataset. The results provide a direct comparison of how well explanations aligned with domain-informed feature sets. We provide FAP, FAR, and FAF1 values for each model under varying \(k\)-values (number of top influential features considered). Different values of \(k\) allow us to examine the model's explanation alignment when considering concise or extensive feature sets from the XAI method.

\begin{table}[htbp]
\centering
\caption{The dataset level explanation evaluation metrics (FAP, FAR, FAF1) for RF, DNN, and CNN-BiLSTM X-IDSs, evaluated across varying top-\(k\) values to assess the alignment of their explanations with domain-informed features.}
\label{tab:dos_cleaned}
\begin{tabular}{lccccccccc}
\toprule
\textbf{Top-\(k\)} & \multicolumn{3}{c}{\textbf{RF}~\cite{arreche2024xai}} & \multicolumn{3}{c}{\textbf{DNN}~\cite{arreche2024xai}} & \multicolumn{3}{c}{\textbf{CNN-BiLSTM}~\cite{sinha2020efficient}} \\
\cmidrule(lr){2-4} \cmidrule(lr){5-7} \cmidrule(lr){8-10} 
 & \textbf{FAP} & \textbf{FAR} & \textbf{FAF1} & \textbf{FAP} & \textbf{FAR} & \textbf{FAF1} & \textbf{FAP} & \textbf{FAR} & \textbf{FAF1} \\
\midrule
5   & 0.09 & 0.04 & 0.06 & 0.30 & 0.18 & 0.23 & 0.17 & 0.09 & 0.12 \\
10  & 0.07 & 0.06 & 0.07 & 0.23 & 0.25 & 0.24 & 0.16 & 0.16 & 0.16 \\
20  & 0.09 & 0.25 & 0.13 & 0.21 & 0.44 & 0.28 & 0.18 & 0.35 & 0.24 \\
40  & 0.11 & 0.63 & 0.19 & 0.17 & 0.70 & 0.27 & 0.20 & 0.82 & 0.32 \\
\bottomrule
\end{tabular}
\end{table}

\subsubsection{Dataset level explanation evaluation results:} 
\begin{figure}[ht]
    \centering
    \begin{subfigure}{0.494\textwidth}
        \centering
        \includegraphics[width=\linewidth]{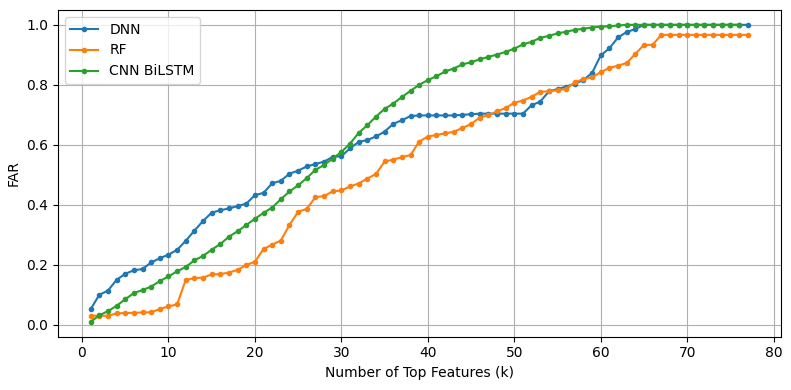}
        \caption{{Dataset level FAR}}
        \label{fig:Dataset_level_FAR}
    \end{subfigure}
    \hfill
    \begin{subfigure}{0.494\textwidth}
        \centering
         \includegraphics[width=1\linewidth]{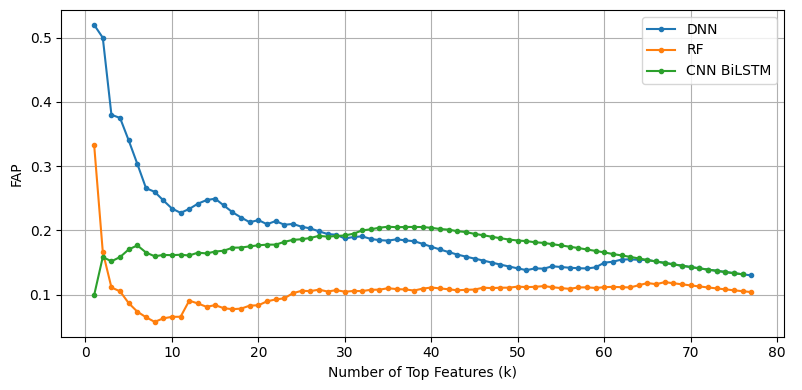}
        \caption{{Dataset level FAP}}
        \label{fig:Dataset_level_FAP}
    \end{subfigure}
    \caption{(a) FAR and (b) FAP at various top-\(k\) cutoffs for DNN, RF, and CNN-BiLSTM X-IDSs. These metrics show how well each X-IDS's top-\(k\) features align with the set of domain-informed features \emph{across the entire dataset}.} 
    \label{fig:Dataset_level_plot}
\end{figure}

We compute the dataset-level evaluation metrics: FAP, FAR, and FAF1, by averaging across all attack classes in the evaluation in order to provide a single overall measure. This measure represents the quality of the explanation for each X-IDS at each \(k\). As shown in Table~\ref{tab:dos_cleaned}, the DNN and CNN-BiLSTM X-IDSs have consistently higher alignment scores across \(k\) values compared to the RF X-IDS. This indicates that the explanations of the deep learning X-IDS (DNN and CNN-BiLSTM) are more aligned with the domain-informed features than the RF X-IDS. 
 For example, at a small cutoff number of features \(k\)= 5, the DNN reaches a FAP of 0.30, where RF is 0.09. This means that 30\% of the top features in DNN are relevant to the attack according to the set of domain-informed features, while only 9\% for the RF features. Similarly, FAR at \(k\)= 5 shows that DNN is quicker in covering more of the domain-informed features (0.18) compared to RF, which covers 0.04 of the expected features. The CNN-BiLSTM also outperforms RF at \(k\)=5 (FAP 0.17, FAR 0.09), but not to the extent of DNN. Furthermore, we notice that the DNN X-IDS provides the highest alignment at smaller \(k\), while the CNN-BiLSTM reaches DNN when \(k\) grows larger, as illustrated in Figure~\ref{fig:Dataset_level_plot}. The DNN achieves the highest FAF1 score at \(k\) = 5 and 10 (0.23 and 0.24), compared to CNN (0.12 and 0.16) and RF (0.06 and 0.07). This indicates that with fewer top influential features from the XAI method, the DNN X-IDS can identify critical domain-informed features that provide a more balanced combination of correctness (FAP) and completeness (FAR). This can provide a practical value as security analysts can utilise quick and accurate insights based on the most influential features in a model prediction. 

\subsubsection{Class level explanation evaluation results:} 
\begin{figure}[ht]
    \centering
    \begin{subfigure}{0.494\textwidth}
        \centering
        \includegraphics[width=\linewidth]{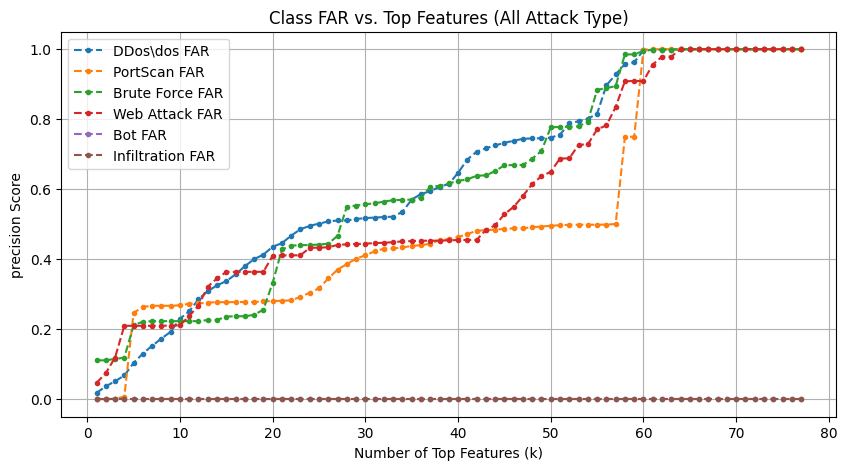}
        \caption{{Class level FAR by attack classes}}
        \label{fig:all_c_FAR}
    \end{subfigure}
    \hfill
    \begin{subfigure}{0.494\textwidth}
        \centering
         \includegraphics[width=1\linewidth]{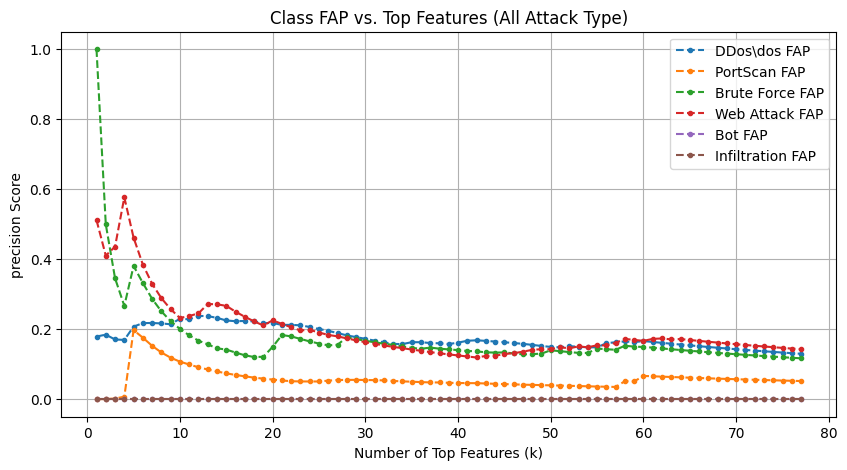}
        \caption{{Class level FAP by attack classes}}
        \label{fig:all_c_FAP}
    \end{subfigure}
    \caption{Class level explanation evaluation metrics across attack types for the DNN model. (a) FAR trends showing how many of the domain-informed features are captured as \(k\) increases. (b) FAP trends showing how many of domain-informed features are captured by the explanations at each \(k\).} 
    \label{fig:multi_attack_curves}
\end{figure}

We also compute the class level FAP, and FAR by averaging the instance-level scores for each attack class. This provides a detailed view of how well each of the explanations of the model aligns with the domain-informed feature sets for individual attack types at different \(k\) values. To better understand how explanation alignment varies across different types of attacks, we visualise class level FAR and FAP with respect to \(k\) across multiple attack classes in the dataset, as illustrated in Figure~\ref{fig:multi_attack_curves}. Each curve corresponds to a different attack class. In Figure~\ref{fig:all_c_FAR}, attack classes such as DDoS/DoS, Brute Force, and Web Attack show improvement in FAR as \(k\) increases. For these attacks, a larger portion of the domain-informed features is captured in the explanation gradually through \(k\). We see that certain attacks reach high FAR with smaller \(k\). For example, the PortScan attack class can capture more domain-informed features at \(k\) = 4 with higher FAR. This indicates that the FAR curve rises quickly for attack classes with a small set of domain-informed features, compared to attack classes with a richer set of domain-informed features, such as DDoD/DoS. In contrast, infiltration and bot attack classes consistently show zero FAR as the predefined domain-informed feature set indicates no features to compare against. In Figure~\ref{fig:all_c_FAP}, we observe that FAP generally decreases as \(k\) increases, which is an expected trend since the comparison of top influential features of explanation at certain \(k\) against the predefined domain-informed feature set will decrease (after exceeding the maximum number of the existing predefined features in the corresponding set). For Web Attack and Brute Force, FAP starts high and drops gradually, which indicate strong alignment at the lower \(k\) values. These trends of different attack classes confirm that our explanation metrics are affected by the specific number of each domain-informed feature per attack type. This variation of alignment scores allows the security analysts to examine which attack classes are well-explained by a model.

\subsubsection{Alignment Trends and Trade-Off Analysis:} To further examine how well the top‑\(k\) features produced by each X-IDS align with the domain-informed feature sets, we further analyse  two aspects: (1) the overall balance of explanation correctness and completeness using the FAF1 metric, and the trade-off between FAP and FAR using a precision–recall style curve. These plots allow us to visually compare how each model performs across different values of 
\(k\). For clarity, we focus on the DDoS/DoS class as a representative example, since it has one of the largest domain-informed feature sets (10 expected features), and the class level view makes it easier to observe how models behave around a known threshold (i.e., number of expected features), especially when visualising FAP and FAR together where \(k\) is implicit. 

\begin{figure}[ht]
    \centering
    \begin{subfigure}{0.494\textwidth}
        \centering
        \includegraphics[width=\linewidth]{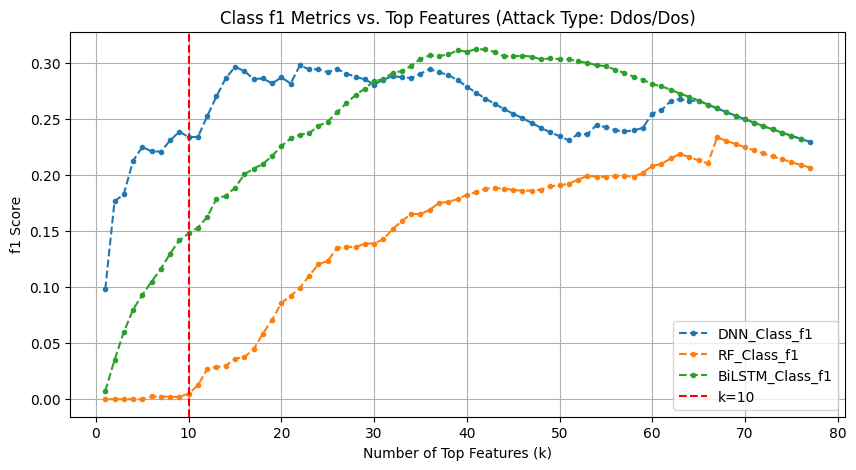}
        \caption{Class Level FAF1 }
        \label{fig:f1_c}
    \end{subfigure}
    \hfill
    \begin{subfigure}{0.494\textwidth}
        \centering
         \includegraphics[width=1\linewidth]{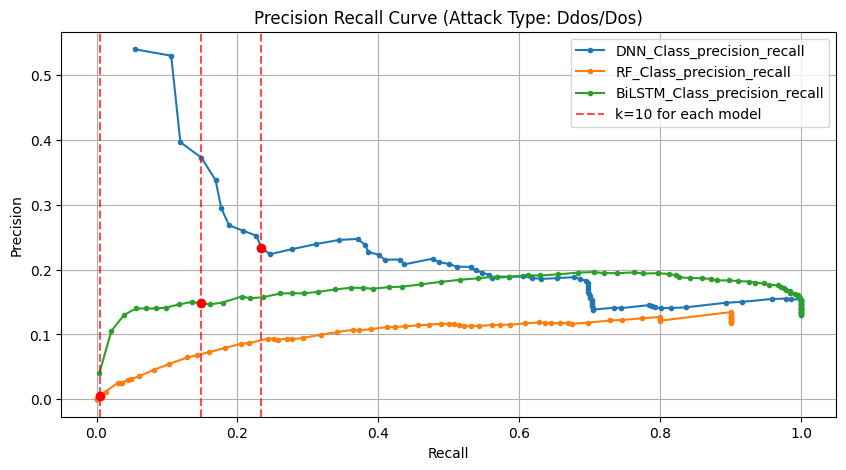}
        \caption{Class Level PR Curve}
        \label{fig:PR_c}
    \end{subfigure}
    \caption{(a) Class Level FAF1 curve illustrating how each model’s top-\(k\) features balance correctness (FAP) and completeness (FAR) with respect to the MITRE-based features. 
    (b) Class Level Feature alignment Precision–Recall (FPR) curve, showing how FAP varies with FAR for different \(k\) values. }
    \label{fig:Both_level_plot}
\end{figure}

\paragraph{FAF1 trend analysis:} Figure~\ref{fig:f1_c} shows the FAF1 curves for the DDoS/DoS class. This plot provides an overall view of alignment quality as \(k\) increases. We observe that the DNN model maintains higher FAF1 values for small \(k\) where it peaks at \(k\)=15. This indicates an optimal balance of correctness and completeness at that point. The FAF1 curve of DNN starts to decline gradually as more irrelevant features are introduced. On the other hand, the CNN-BiLSTM curve for FAF1 gradually exceeds DNN, however, this happens slowly at a larger number of \(k\). The RF curve for FAF1 reflects poor balance as it remains the lowest across all \(k\) values. This indicates that it does not capture the MITRE-based features effectively at any point. Overall, FAF1 plot offers a holistic view that helps to illustrate the balance of both completeness (FAR) and correctness (FAP) and highlight the overall quality of alignment for each model.

\paragraph{FAP–FAR trade-Off analysis:} Figure~\ref{fig:PR_c} shows another perspective that represents a FAP against FAR for each model. It provides a precision-recall style view of the alignment by directly plotting FAP (precision) against FAR (recall) for the DDoS/DoS class. Each curve represents the values of FAP and FAR values at each \(k\) that is implicitly shown by the number of points. The FAP-FAR comparison curve highlights how each model balances FAP and FAR across various values of \(k\), with vertical lines highlight particular cuttoffs at \(k\) = 10, which corresponds to the number of domain-informed features for the evaluated attack class. The DNN model, for example, begins with a high FAP at a modest FAR. This indicates that more expected features are retrieved and covered in the early top features generated by the most influential features. The FAP-FAR comparison curve is important because it highlights the potential trade-off between FAP and FAR for each model in a single view. Compared to the separated plots of FAP and FAR over \(k\), the combined curve allows us to assess how the model is efficiently retrieving domain-informed features with merely the values of FAP and FAR. As \(k\) values are not explicitly shown, vertical reference lines can help indicate key cutoff points for comparing model behaviour at meaningful thresholds. This curve offers an intuitive view of the trade-off between FAP and FAR to identify how well a model retrieves domain-informed features.

\subsection{Key insights and implications}
The above results demonstrate that explanations for X-IDSs can be quantitatively evaluated against domain knowledge using our evaluation metrics. The DNN and CNN-BiLSTM X-IDSs show a better alignment with MITRE-informed features than the RF. This indicates that the ability of an X-IDS to produce explanations aligned with domain knowledge is a critical aspect of explanation quality, and it is not equal between models. For security analysts, this means that an X-IDS with higher alignment can produce more useful and interpretable alerts. This alignment can speed up the understanding of an alert, in order to quickly respond to it. 

An important factor influencing explanation quality is the value to consider for the top-\(k\) features. Our evaluation across multiple \(k\) shows how the quality of each explanation differs. The patterns of different models suggest that explanations can offer meaningful insight with relatively few features when those features are highly aligned with domain knowledge. At lower \(k\), a higher FAP indicates that the most influential features included are highly relevant. Similarly, a higher FAR value shows more complete coverage of the expected feature set. In particular, this perspective is best observed through the FAF1 curve, which reveals the effectiveness of explanation's quality across different cutoffs to determine the most appropriate \(k\) value. This has a practical use when deciding whether we aim to show analysts a very concise explanation, or we can afford more features to be shown. Thus, we can choose \(k\) based on the trade-off presented in FAF1 and the FAP-FAR comparison curve. 

In addition to FAF1, we plot FAP-FAR comparison curves across different \(k\) to show the trade-off between explanation correctness (FAP) and completeness (FAR). Unlike the standard Precision-Recall (PR) curve that relies on a continuous probability threshold, our version depends on discrete \(k\) to calculate FAR and FAP values. Consequently, when we plot multiple models together, the progression along \(k\) may differ in how quickly or gradually models retrieve relevant features. As a result, the comparison can be visually misleading if one model adds relevant features earlier than another. Although both axes still run from 0 to 1 for FAP and FAR in the PR-curve, the stepwise progression of each model can be visually confusing. To address this challenge, we mark key \(k\) values, such as when \(k\) equals the number of expected features for a specific attack label (e.g., \(k\) equals 10 for DDos/Dos), so the relative performance of different models at meaningful thresholds becomes easier to interpret.

Another insight relates to the quality of the domain-informed feature sets themselves. It is important to evaluate the results in the context of the quality and completeness of the predefined domain-informed feature sets. However, in cases such as the infiltration and bot attacks curves, we observed low alignment scores. Such low alignment scores suggest that the domain-informed features derived from expert knowledge may not fully capture relevant indicators or may not generalise well to the dataset used. Due to this mismatch, we find that the predefined domain-informed feature sets may require additional fine-tuning to align with features in the selected dataset. Consequently, the score of the alignment metrics can be affected. Therefore, it needs to include a user feedback loop in order to adjust the feature sets when necessary. 

Finally, our current evaluation metrics aggregate FAP and FAR using the arithmetic mean. Consequently, averaging the values provides a single and intuitive number that shows how well the performance is across all instances. Although the arithmetic mean is straightforward and easy to interpret, it does treat each instance as equally important. This makes it vulnerable to outliers in a skewed distribution. In such cases, alternative aggregation techniques, such as median, weighted, or trimmed averages, can offer more robust and reliable evaluation, especially when the class or dataset level distributions contain outliers. 

Overall, our findings show that the proposed explanation evaluation metrics (FAP, FAR, FAF1) can serve in several purposes. First, they effectively differentiate between X-IDSs based on how well their explanations align with domain knowledge. This differentiation makes it easier to identify which X-IDS produce more meaningful outputs. Second, by analysing how the explanations quality are changing across \(k\), the metrics offer insights on how many top-\(k\) features should be shown to analysts. Finally, when the metrics scores are constantly low for certain attack classes, these scores can indicate that the domain-informed feature sets might need to be revisited or refined. Therefore, these metrics can be valuable in assessing the explanations' quality when developing X-IDSs, as well as, improving how explanations are presented to security analysts when deployed in practical settings.

\section{Conclusion}

In this paper, we present three novel metrics to evaluate the explanations generated by X-IDSs with domain-specific knowledge in the context of intrusion detection systems. We analyse our three explanation evaluation metrics for a popular XAI technique (i.e., SHAP). These three evaluation metrics are Feature Alignment Precision (FAP), which quantifies the correctness of the output of the XAI method based on defined indicators from domain-specific knowledge, Feature Alignment Recall (FAR), which quantifies how well the most important features capture all defined domain knowledge, and Feature Alignment F1 (FAF1), which quantifies the harmonic means between FAP and FAR. Our metrics enable the explanation evaluation at three different levels: instance, class, and dataset levels in order to capture the quality of the explanation, respectively, for individual predictions, across specific attack types, and obtain an overall measure of interpretability.

We applied our metrics to explanations generated for three X-IDS (Random Forest, DNN, CNN-BiLSTM) trained on a balanced subset of the CICIDS2017 dataset. The experimental results demonstrated how our metrics provide a richer and more actionable view of explanation quality. Specifically, our findings highlight that X-IDS can offer higher alignment with expected features at lower values of \(k\), which makes their explanations more suitable for operational use. 

As future work, we plan to investigate alternative aggregation strategies, such as weighted averages, as we recognise that our approach assumes every instance contributes equally and could be heavily influenced by outliers. Additionally, we aim to apply the proposed evaluation metrics across various machine learning models and explanation methods to better understand their alignment with domain knowledge.

\end{document}